\documentclass[letterpaper, 10 pt, conference]{IEEEtran} 
 \pdfoutput=1
\usepackage{pslatex}  
\usepackage{graphicx}  
\usepackage{color}     
\usepackage{amsmath} 
\usepackage{multirow}
\usepackage{geometry, url}
\usepackage{pstricks}
\usepackage{pstricks-add}
\usepackage{pst-plot}
\usepackage{pst-grad}
\usepackage{subfigure}
\usepackage{amssymb}  
\usepackage{epstopdf,subfigure}

\usepackage{amsmath}
\usepackage{framed, url}

\IEEEoverridecommandlockouts

\begin{document}           

\rmfamily 
\upshape 





\title{\LARGE \bf On the Statistics and Predictability of Go-Arounds}

\author{Maxime Gariel, Kevin Spieser, and Emilio Frazzoli\thanks{This research is partially supported by NASA grant \# NNX08AY52A.}
\thanks{
{\tt\small \{mgariel,kspieser,frazzoli\}@mit.edu}}
}
\addtolength{\baselineskip}{-0.05\baselineskip} 
\maketitle

\begin{abstract}
This paper takes an empirical approach to identify operational factors at busy airports that may predate go-around maneuvers. Using four years of data from San Francisco International Airport, we begin our investigation with a statistical approach to investigate which features of airborne, ground operations (e.g., number of inbound aircraft, number of aircraft taxiing from gate, etc.) or weather are most likely to fluctuate, relative to nominal operations, in the minutes immediately preceding a missed approach. We analyze these findings both in terms of their implication on current airport operations and discuss how the antecedent factors may affect NextGen. Finally, as a means to assist air traffic controllers, we draw upon techniques from the machine learning community to develop a preliminary alert system for go-around prediction.      
     
\end{abstract}


\section{Introduction}
A missed approach or go-around (GA) occurs when an aircraft aborts its landing and is forced, instead, to land on a subsequent approach. It is tempting to speculate that lack of visibility at decision height or pilot error, that is a pilot's inability to safely land the aircraft in a given situation, are a leading cause of GAs. Indeed, low ceiling and visibility increase the potential for missed-approach and the workload for pilots and controllers \cite{JPDO2010Weather}. Nevertheless, as it is shown in this paper,  weather only accounts for a small fraction of the total number of go-arounds at SFO. However, interviews with air-traffic controllers during field visits to Logan International Airport in Boston and Laguardia Airport in New York refute these claims; rather, the controller's testimony suggests operational errors, such as a runway incursion, late runway departure from an aircraft taking off or holding position line violation are the primary causes of GAs. 

To increase airport throughput, NextGen's high-density operations~\cite{JPDOConops3.2} are projected at more airport's than today's class B airports. High-density operations require high performance procedures and aircraft equipage to enable Closely Spaced Parallel Approaches with delegated separation procedures. Despite the increased automation, the technologies, there will always be errors or unexpected events leading to missed approach. Avionics for de-conflicted missed approaches for converging is still in the roadmap of NextGen but has not been addressed yet \cite{JPDO2008Avionics}. To take full advantage for these high throughput operations, the number of missed approaches needs to be minimized, and therefore a thorough understanding of the factors that lead to missed approach is  necessary.

Go-arounds intensify the workload of air traffic controllers, as landing sequences must be amended to accommodate the aircraft that failed to land \cite{boursier2006airborne}. On a related note, the requisite revamping schedules taxes an airport system that maintains high-safety standards largely through comprehensive planning and delegation, go-arounds are also undesirable from a safety perspective \cite{Thiel10}. Finally, go-arounds are costly for airlines, both in terms of the added fuel cost and the logistic delays absorbed from spending extra time airborne \cite{Robertson10}. Finally, as go-arounds require personnel to quickly shift to quickly invoke  
 
 This report is an exploratory study aimed at addressing the following questions i.) can we identify factors that lead to GAs and ii.) if so, can these causes be identified in real-world data sets from major airports to predict GAs?   If the factors causing GAs can be identified, then mitigation action can be taken in order to reduce the number of GAs without impacting airport throughput.

The remainder of this report is structured as follows. Section~\ref{sec:data} describes the datasets used to construct our corpus of flight data. Section~\ref{sec:stats} provides a statistical analysis of conditions that may be closely associated with GAs and Section~\ref{sec:prediction} explores the idea of using these findings as a means to predict GAs. Conclusions are presented in Section~\ref{sec:conclusions}. 

\section{Data presentation and selection}\label{sec:data}
In an attempt to identify factors that precipitate a GA, data was collected from three complimentary datasets over a four-year period spanning January 2006 to December 2009. Following cleaning and synchronization, the data was combined to create a single master dataset.
\subsection{Data presentation}
\begin{figure}
\centering
\includegraphics[width = 0.3\textwidth]{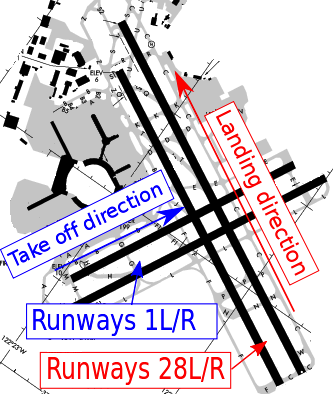}
\caption{Simplified SFO diagram and selected runway configuration}\label{fig:SFO}
\end{figure}

The datasets used to build our corpus of flight data are described below and the fields extracted from the datasets are presented in Table~\ref{tab:airspace}. The values of the fields were sampled for every minute in time during the four years of study. If a GA occurs during occurs within the one minute sample window we refer to the sample as a GA sample. Otherwise, the sample is referred to as a nominal sample. In the event a portion of the data associated with a particular sample of interest was absent, the sample was removed from the data set in order to ensure uniformity among dataset entries. The study focused on the periods of higher traffic density, that is 7:00 to 23:00 local time.\\
\noindent\textbf{Airspace data:} This dataset was provided by the San Francisco Noise Abatement Office ~\cite{SFOnoise}. It contains data for all of the flights recorded by the secondary radar located at Oakland International Airport (OAK). 
In the data set, it appears the range of the radar was increased from 45 to 60 NM over the four years of interest and to ensure consistency over the 4 year period only the data in a radius smaller than 45 NM was kept. For each flight, the dataset contains the aircraft's 4-D trajectory as well as metadata such as flight identification, origin airport, destination airport, etc. We kept only flights departing or arriving at one of the three largest airports in the Bay Area, that is San-Francisco International Airport (SFO), Oakland International Airport (OAK) and San Jose International Airport (SJC). \\
\noindent\textbf{Ground data:} The ground data was extracted from the Aviation System Performance Metric (ASPM) flight database. This database contains a record of both the scheduled time and actual time for pullback from gate, takeoff, and landing for each aircraft at each major airport in the United States. This data is available for download from the Federal Aviation Administration (FAA) website
~\cite{FAAaspm}. To compliment the airspace fields, the relevant fields for all flights taking off or landing at SFO were extracted from the ASPM dataset.\\
\noindent\textbf{Weather and runway data:} The weather information and the runway configuration for SFO was extracted from the ASPM Airport database. This database is also available on the FAA website~\cite{FAAaspm}. The database includes the weather (visibility, cloud ceiling, wind speed and direction) as well as the runway configuration in use. Figure~\ref{fig:SFO} depicts the layout of SFO. In the configuration illustrated, aircraft take off from runways 1R-L (parallel runways), and land on runway 28R-L (also parallel runways); this is the configuration in use approximately 80 percent of the time. Among weather-related fields, the temperature was not always available at each time instant. In these cases, the temperature measurement used was obtained by interpolation neighboring entries. 
\begin{table}[!htbp]
	\centering
	\caption{List of variables generated }\label{tab:airspace}
\begin{center}
\begin{scriptsize}
\vspace*{-15pt}
\begin{tabular}{|l|l|}
\hline
Index \#& Fields Name: ``Airspace data'' \\
 \hline
 1&  Time of the day\\
 %
 %
 2-5&  \# of ac, SFO  inbound, current, average 5, 10, 15 min\\
 6-8& \# of ac, SFO inbound, variation 5, 10, 15 min\\
 %
 %
 9-12& \# of ac, SFO outbound, current, average 5, 10, 15 min\\
 13-15& ac, SFO outbound, variation 5, 10, 15 min\\
 %
 %
 16-29& \# Same as 2-15 for OAK.\\
 %
 %
 %
 %
 %
 %
30-43& \# Same as 2-15 for SJC \\
%
%
%
%
%
44-46& Landing rate at SFO (ac/min) 5, 10, 15 min\\
47-49& Time elapsed to land the previous 4, 8, 12 ac at SFO\\
50-52& Departure rate at SFO (ac/min) 5, 10, 15 min\\
53-55& Time elapsed to takeoff the previous 4, 8, 12 ac\\
56-58& Landing rate at OAK (ac/min) 5, 10, 15 min\\
59-61& Time elapsed to land previous 4, 8, 12 ac at OAK\\
62-64& Departure rate at OAK 5, 10, 15\\
65-67& Time elapsed to takeoff the previous 4, 8, 12 ac at OAK\\
68-70& Landing rate at SJC (ac/min) 5, 10, 15 min\\
71-73& Time elapsed to land the previous 4, 8, 12 ac at SJC\\
74-76& Departure rate at SJC (ac/min) 5, 10, 15 min\\
77-79& Time elapsed to takeoff the previous 4, 8, 12 ac at SJC\\
80-88& \# of small, large, heavy, ac landing at SFO\\
& in past 5, 10, 15 min\\
%
%
%
%
%
89-97& \# of small, large, heavy, ac taking-off from SFO\\
& in past 5, 10, 15 min\\

\hline
\hline
Index& Fields Name: ``Ground data''\\
\hline
%
98-101& \# of ac taxiing in, current, average 5, 10, 15 min\\
102-104& \# of ac taxiing in, variation 5, 10, 15 min\\
%
%
105-108& \# of ac taxiing out, current average 5, 10, 15 min\\
109-111& \# of ac taxiing out, variation 5, 10, 15 min\\
112& Total estimated departure delay\\
113-117& \# of ac out delayed $>$ 0, 10, 20, 30, 45 min\\
118&  Average delay by aircraft, out\\
119& Total estimated delay from schedule, arrivals\\
120-124& \# of ac delayed in $>$ 0, 10, 20, 30, 45 min\\
125& Average delay by aircraft, in\\
\hline
\hline
Index& Fields Name: ``Weather data''\\
\hline

126& 1 for visual MC and 0 for instrument MC\\
127-129& Ceiling, Visibility, Temperature\\
%
%
%
130-133& Wind Angle, windspeed, headwind, crosswind\\
%
%
134& Number of Runway(s) used for landing\\
135& Number of Runway(s) used for take offs\\
%
%
\hline
\end{tabular}
\end{scriptsize}
\end{center}
\label{default}
\end{table}
\subsection{GA detection and selection}
\begin{figure}[ht]
 \centering
 \subfigure[Trajectory]{\includegraphics[width = 0.48\linewidth]{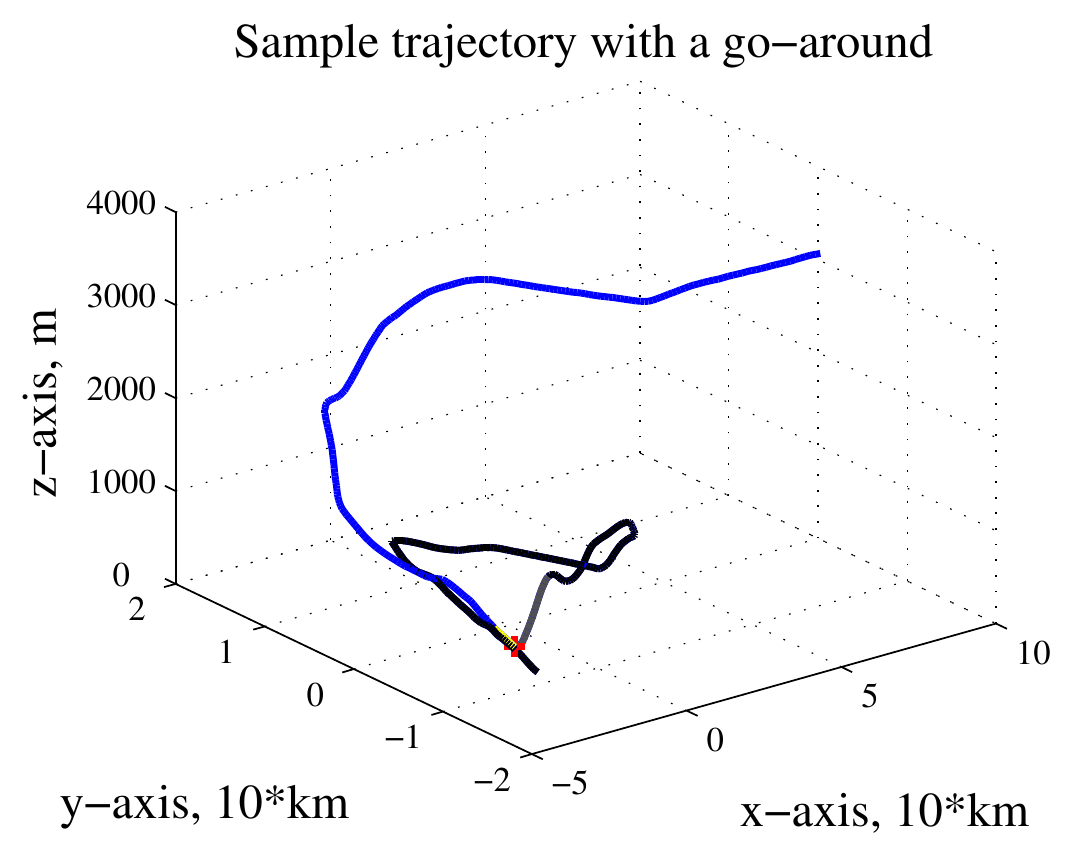}}
 \subfigure[Vertical profile]{\includegraphics[width = 0.48\linewidth]{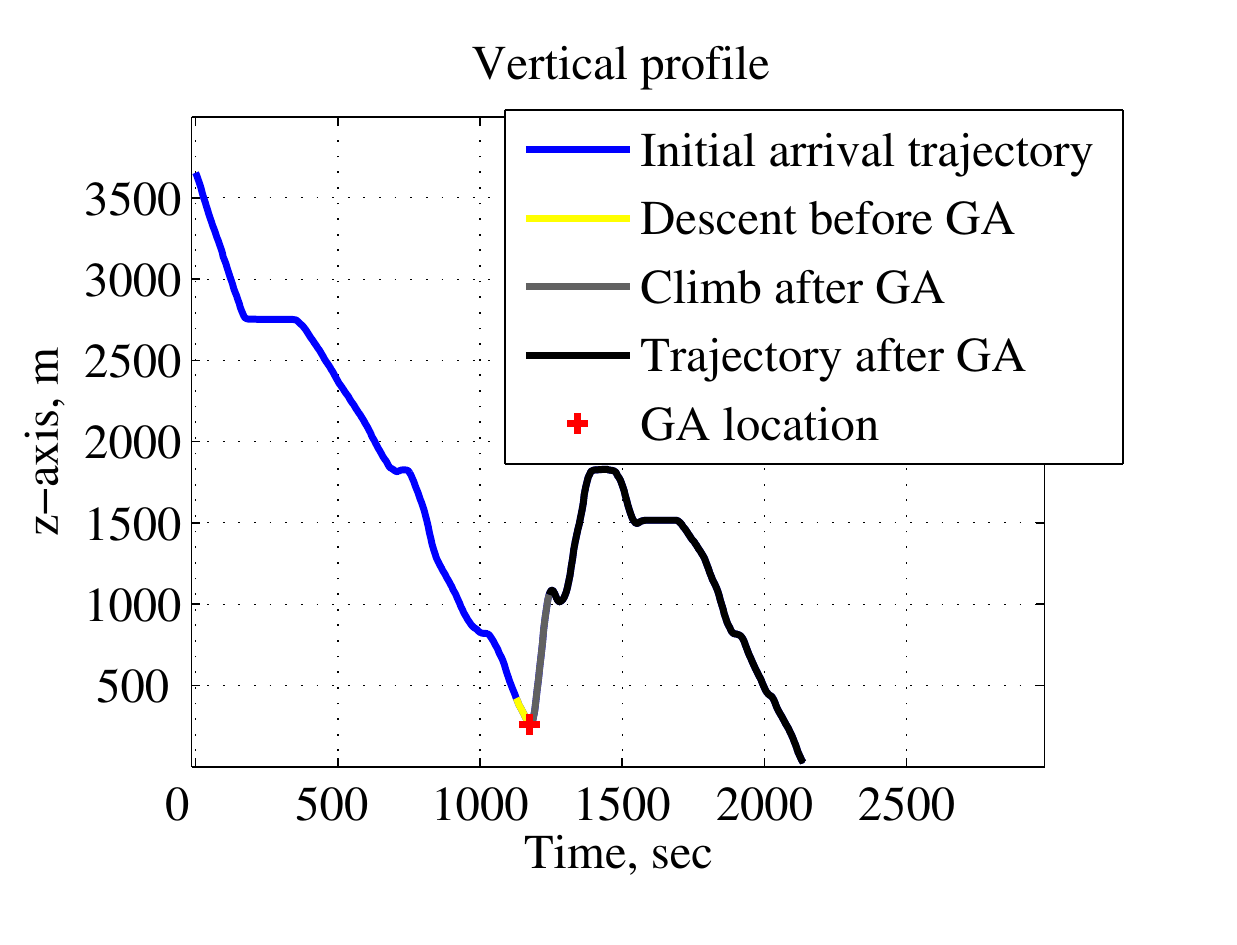}}
 \caption{Sample landing trajectory containing a GA}\label{fig:goAroundExample}
 \end{figure}

To facilitate our investigation of GAs, we assembled a corpus of samples, each of which is one of two types: i.) samples of the airport state during nominal operations in which no GAs occur and ii.) samples of the airport state during a window in which a GA does occur. The following rule was used to label a flight as containing a GA: a flight contains a GA if during the plane's terminal flight phase, the plane's altitude increases for fifteen consecutive measurements following a period in which the plane descended for at least ten consecutive radar measurements. At the sample rate at which measurements were taken, this corresponds to approximately 70 seconds of continuous increase in altitude, following 45 seconds of continuous descent. This criterion identified nearly all GAs and was discerning enough to exclude the trajectories of helicopters and short-haul flights not associated with GAs. This method of detecting GAs was validated through manual verification on a large sample of trajectories. For example, Figure \ref{fig:goAroundExample} shows a sample landing trajectory containing a GA. The blue line shows the portion of the trajectory prior to the GA. The yellow segment corresponds to the 45 seconds of descent preceding the GA. The instant at which the GA is initiated is indicated with a red cross. Finally, the grey line corresponds to the period of at least 70 seconds of climb following the GA. The remainder of the trajectory, including the eventual landing, is shown in black. 

\begin{figure}[ht]
 \centering
 \includegraphics[width = 0.48\linewidth]{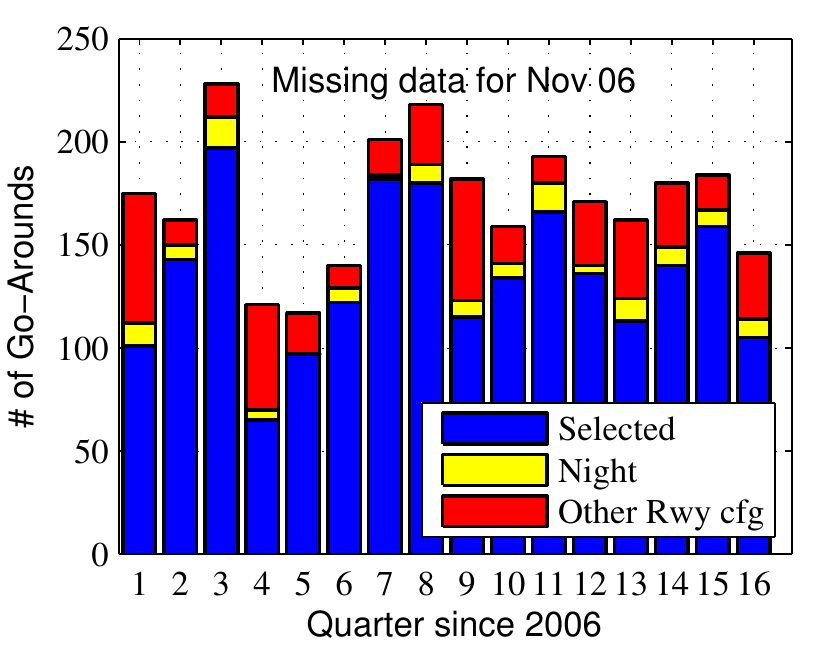}
 \caption{Distribution of GA over time}\label{fig:GATimeHistory}
 \end{figure}

   Figure \ref{fig:GATimeHistory} shows the number of GA gathered from the data, by quarter, starting in 2006. The blue corresponds to the corpus of GA selected for this study, the yellow corresponds to the GA occurring at night (23:00 to 7:00), and in red, the GA occuring on other runway configurations. Due to missing data, these numbers do not reflect the exact count of GA at SFO. 
 \vspace{-10pt}

\section{Parameter analysis}\label{sec:stats}
This section presents a statistical analysis of the distribution of the different variable features for both nominal and GA flights. The nominal samples used in this section consist of 120,000 samples that were randomly taken from points in time no less than 15 minutes in time away from a flight that performs a GA. The GA corpus contains all 2155 GA samples.
For all figures presented in this section, the values corresponding to the GA flights are shown in red, values corresponding to ``nominal" flights are shown in blue. The following discussion highlights operational factors that we found interesting, either because the data showed a significant difference between the nominal and GA sample distributions, or because there was remarkably little difference between the two distributions.   
\subsubsection{Time of day}
\begin{figure}[!ht]
\centering
\includegraphics[width=0.48 \linewidth]{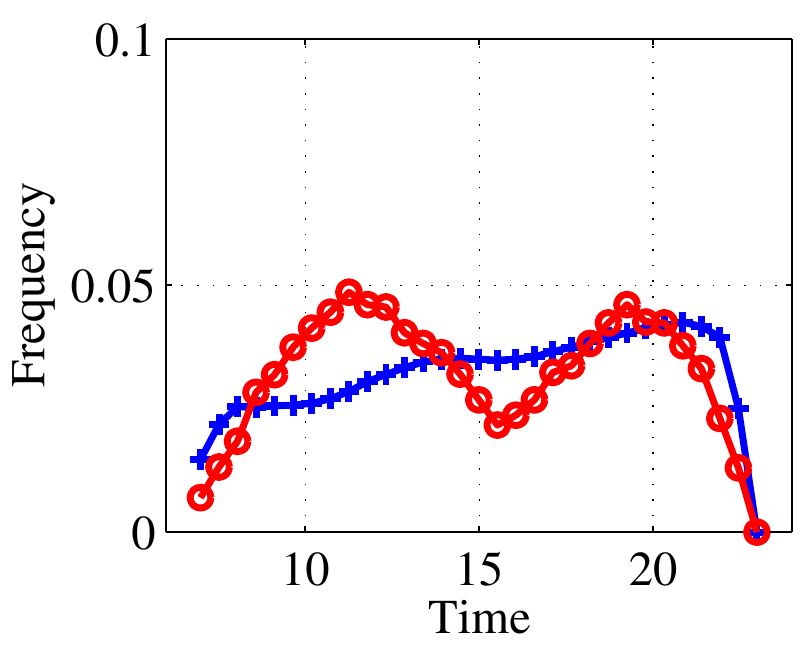}
\caption{Distribution of Time for the nominal observations and the GA}\label{fig:time}
\end{figure}

Figure~\ref{fig:time} presents the sample distributions arranged by time of the day from 7:00 to 23:00 local time. The nominal distribution is not uniform on account of the runway configuration used for study and missing data; on some days, data for the morning was missing for unknown reasons. It appears that there are two peaks where GAs occur more frequently: from 9:00 to 14:00 and then again from 19:00 to 21:00. In the interim, the frequency of GAs achieves a minimum near 15:30. 
%
%
\subsubsection{Number of aircraft inbound for SFO}
\begin{figure}[!ht]
\subfigure{\includegraphics[width=0.48 \linewidth]{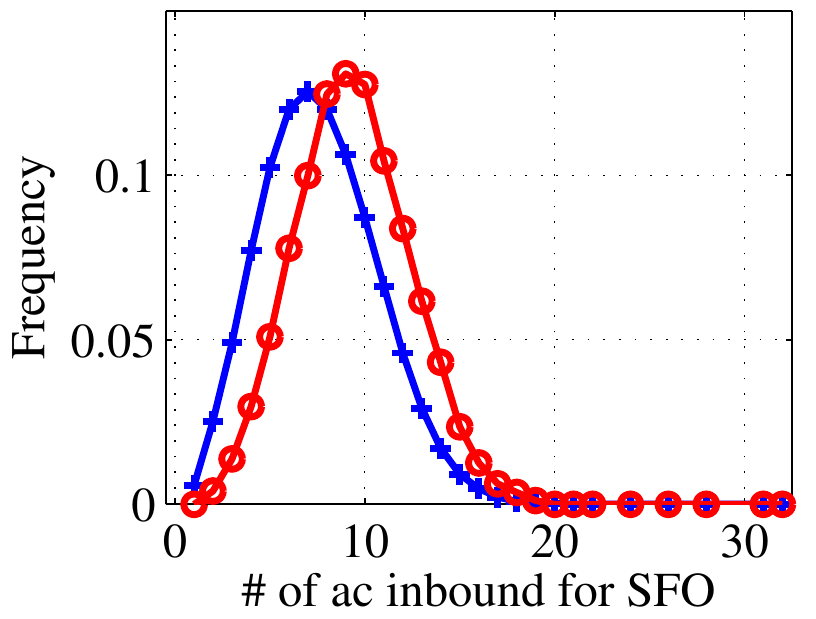}\label{fig:inboundSFO1}}
\subfigure{\includegraphics[width=0.48 \linewidth]{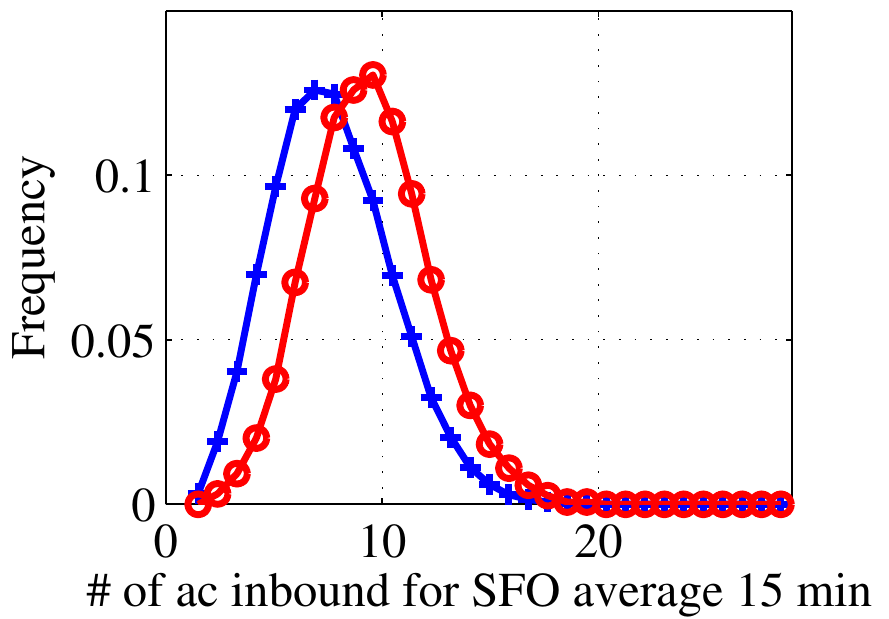}\label{fig:inboundSFO2}}
\subfigure{\includegraphics[width=0.48 \linewidth]{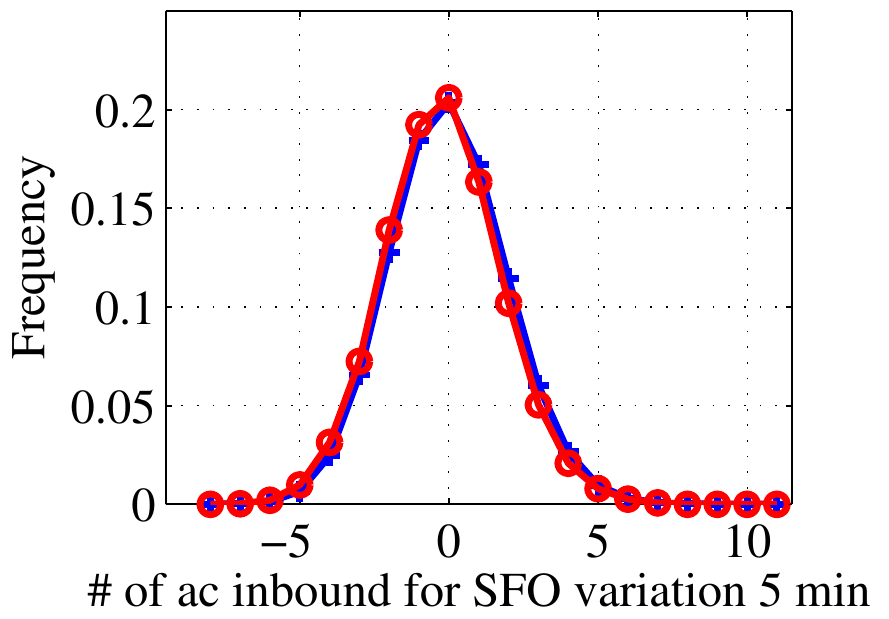}\label{fig:inboundSFO3}}
\subfigure{\includegraphics[width=0.48 \linewidth]{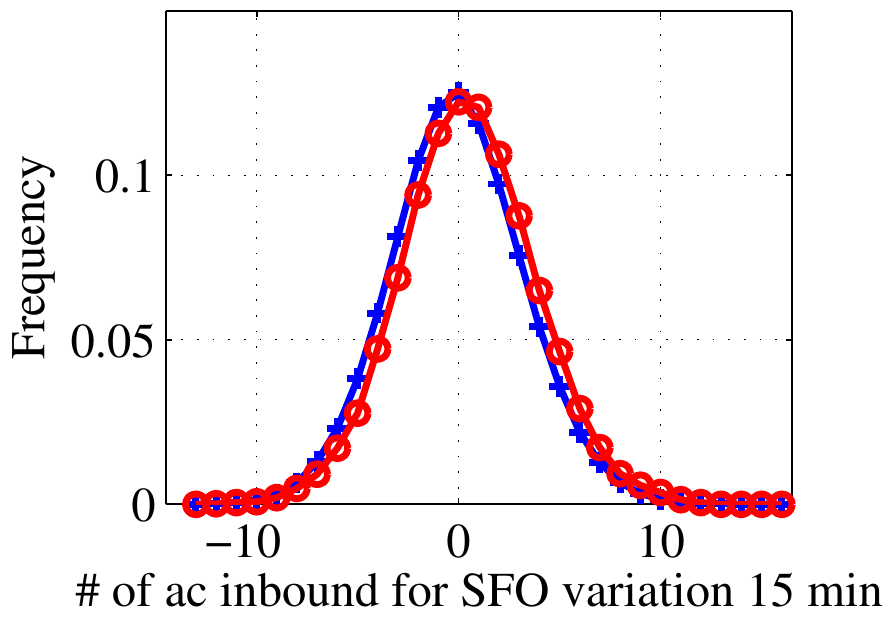}\label{fig:inboundSFO4}}
\caption{Analysis of the number of aircraft incoming to  SFO}\label{fig:inboundSFO}
\end{figure}

Figure~\ref{fig:inboundSFO} shows the distribution corresponding to the number of airborne flights inbound for SFO present in the terminal airspace. Figure~\ref{fig:inboundSFO1} shows the number of aircraft at the time the sample is taken, Figure~\ref{fig:inboundSFO2} shows the average number of aircraft during the 15 minutes preceding the sample. Intuitively, these statistics capture a measure of an air traffic controller's current and recent activity level, respectively. The GA and nominal distributions are similar, but there is a visible shift in the mean; the mean of the GA distribution is approximatively 3 aircraft larger than that of the nominal distribution. 
 Figures~\ref{fig:inboundSFO3} and~\ref{fig:inboundSFO4} present the distributions for nominal and GA flights in the difference between the number of aircraft in the system at present and the number of aircraft 5 and 15 minutes ago, respectively. The 5 minute variation does not illustrate a significant difference between distributions. The 15 minute distributions suggests that the distribution of GAs is shifted slightly to the right relative to the associated nominal distribution. 
\subsubsection{Landing/takeoff rates at SFO and aircraft types}
\begin{figure}[!ht]
\subfigure{\includegraphics[width=0.48 \linewidth]{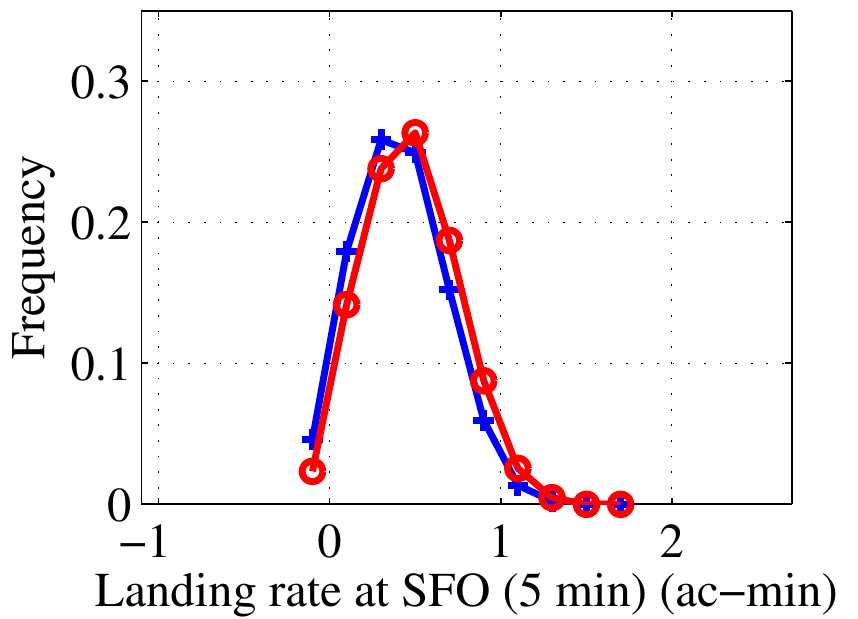}\label{fig:rateSFO1}}
\subfigure{\includegraphics[width=0.48 \linewidth]{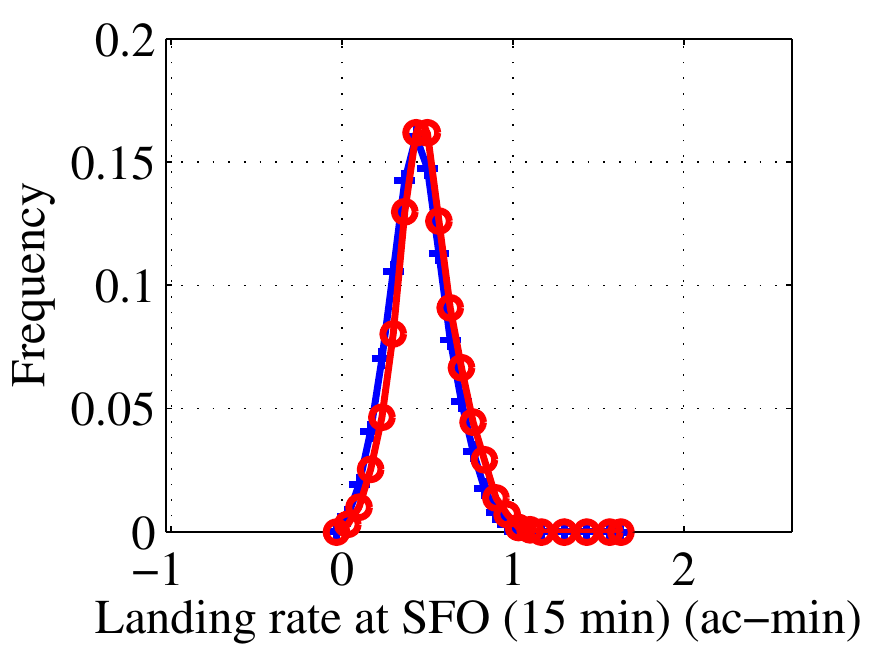}\label{fig:rateSFO2}}
\subfigure{\includegraphics[width=0.48 \linewidth]{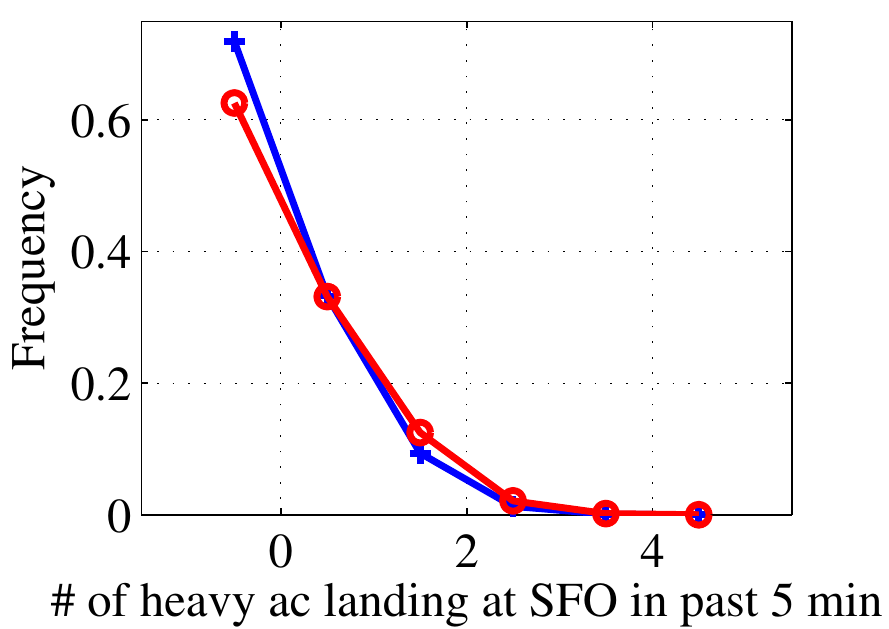}\label{fig:heavySFO3}}
\subfigure{\includegraphics[width=0.48 \linewidth]{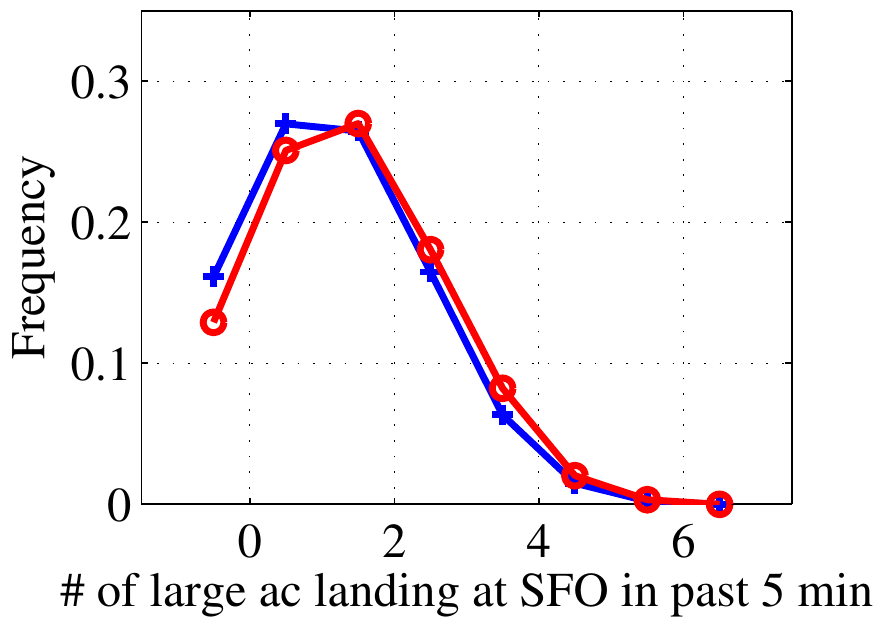}\label{fig:largeSFO4}}
\caption{Analysis landing/takeoff rates at SFO and aircraft types}\label{fig:ratesSFO}
\end{figure}

Figure~\ref{fig:ratesSFO} presents an analysis of the landing rates (number of aircraft landing per minute) over the past 5 and 15 minutes (Figures~\ref{fig:rateSFO1} and \ref{fig:rateSFO2}) as well as the number of heavy and large aircraft landing over the past 5 minutes (Figures~\ref{fig:heavySFO3} and \ref{fig:largeSFO4}). It appears that a higher rate of landing increases the likelihood of a GA, but not in a very important manner. Moreover, it appears that higher numbers of large and heavy aircraft also tend to increase the likelihood of a GA occurring. Although omitted here, the number of small aircraft landing in the past 5, 10 and 15 minutes displayed no significant difference between nominal and GA distributions. Also omitted, for similar reasons, are the distributions corresponding to the takeoff rate at SFO.
%
%
\subsubsection{Number of aircraft outbound from SFO}
\begin{figure}[!ht]
\subfigure{\includegraphics[width=0.48 \linewidth]{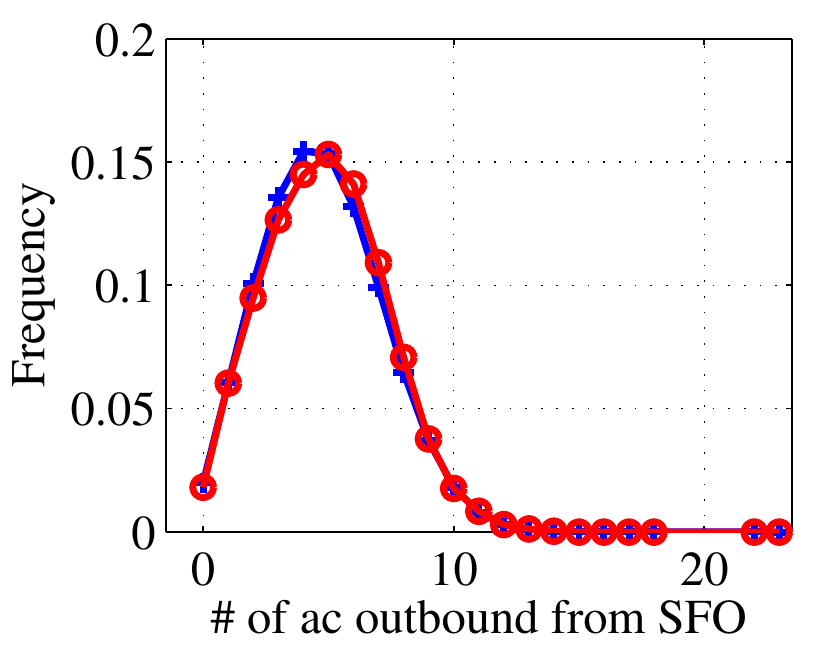}\label{fig:outboundSFO1}}
\subfigure{\includegraphics[width=0.48 \linewidth]{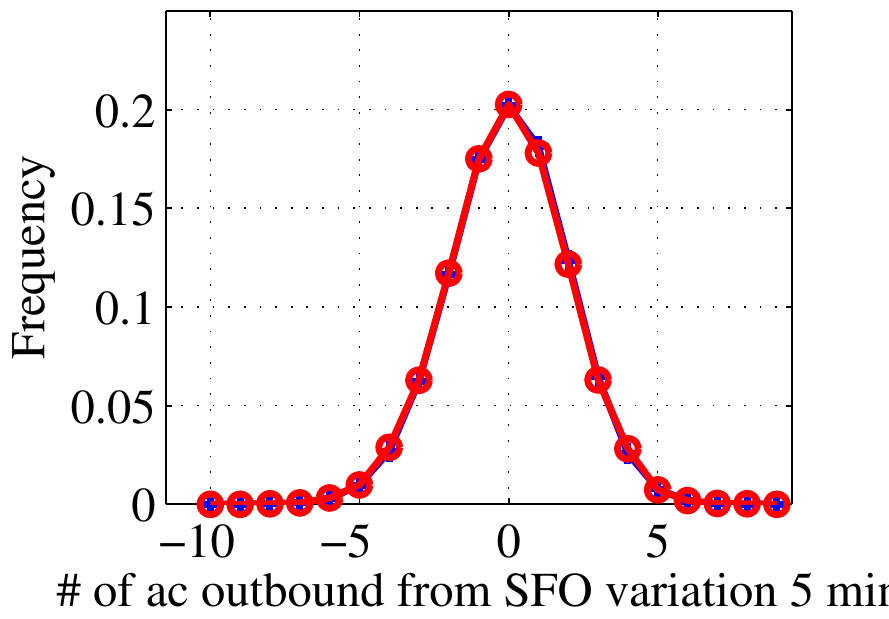}\label{fig:outboundSFO2}}
\caption{Analysis of the number of aircraft outbound from SFO}\label{fig:outboundSFO}
\end{figure}

Figure~\ref{fig:outboundSFO} depicts the distributions associated with the number of aircraft outbound from SFO, and the variation over 5 minutes. There is no significant difference between the data corresponding to GAs and nominal flights. We omit the associated average and variation plots for 5, 10, and 15 minutes as there are no significant differences the nominal and GA distributions. The outbound traffic does not appear to have a statistical impact on the occurrence of GAs.
%
\subsubsection{Number of aircraft inbound for OAK - Landing rate}
\begin{figure}[!ht]
\subfigure{\includegraphics[width=0.48 \linewidth]{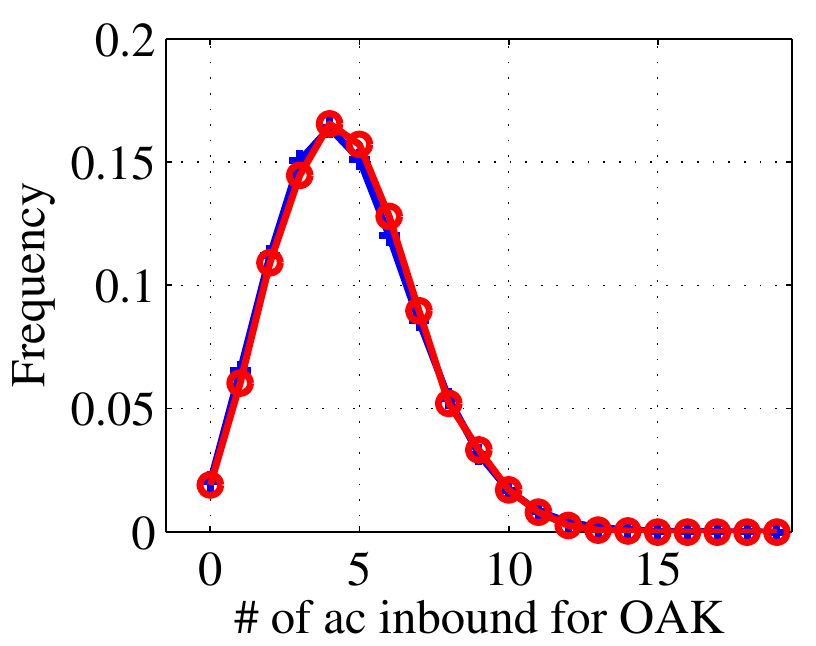}\label{fig:inboundOAK1}}
\subfigure{\includegraphics[width=0.48 \linewidth]{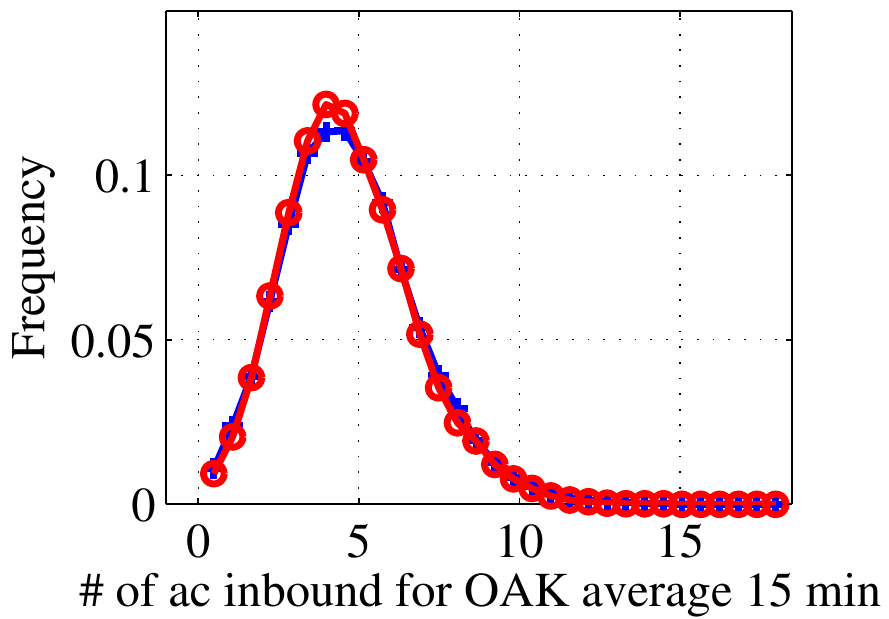}\label{fig:inboundOAK2}}
\subfigure{\includegraphics[width=0.48 \linewidth]{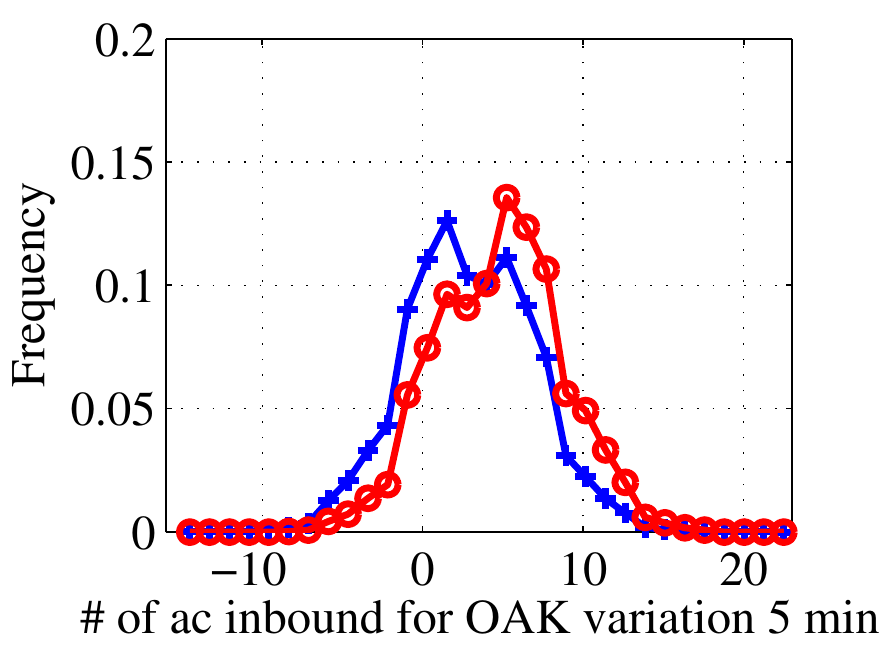}\label{fig:inboundOAK3}}
\subfigure{\includegraphics[width=0.48 \linewidth]{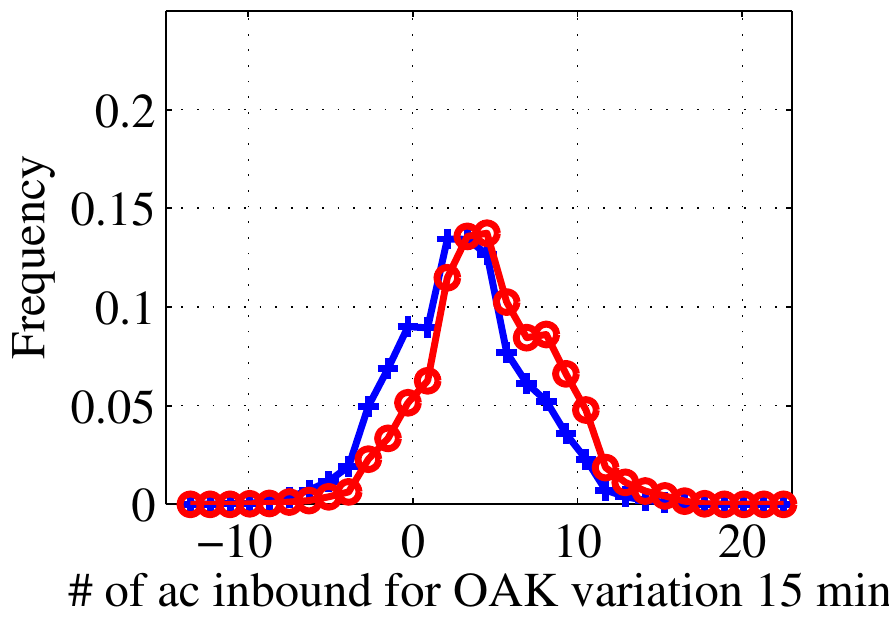}\label{fig:inboundOAK4}}
\caption{Analysis of the number of aircraft incoming to OAK}\label{fig:inboundOAK}
\end{figure}

Figure~\ref{fig:inboundOAK} presents the distribution of the number of flights present in the terminal airspace (in the air) and inbound for OAK. Figure~\ref{fig:inboundOAK1} shows the number of aircraft at the time the sample is taken. Figure~\ref{fig:inboundOAK2} shows the average number of aircraft during the 15 minutes preceding the time of the sample. These measures  reflect the current activity of the controllers and their workload over the past 15 minutes. The nominal and GA distributions do not differ significantly, meaning the number of aircraft inbound for OAK does not appear to have a statistical impact on the occurrence of GAs at SFO.
 Figures~\ref{fig:inboundOAK3} and~\ref{fig:inboundOAK4} present the difference between the number of aircraft simultaneously present at the time of the sample and the number in the system 5 and 15 minutes in the past, respectively. The GA distribution is shifted slightly to the right of the nominal distribution, suggesting GAs occur more frequently when there is an increase in the number of aircraft inbound for OAK during the preceding minutes. Note that the plots are not centered at 0, suggesting a correlation between the runway configuration used at SFO and changes in the traffic volume inbound for OAK.
The landing rates at OAK over the preceding 5, 10 and 15 minutes are not presented, since they do not imply any significant statistical impact on GAs at SFO.
%
\subsubsection{Number of aircraft outbound from OAK - Takeoff rate}
Distributions in the number of aircraft outbound from OAK, the variation in the number of aircraft outbound as well as the takeoff rates are not presented; they present no significant difference between the data corresponding to the GA and nominal samples. 
%
%
%
\subsubsection{Number of aircraft  inbound to/outbound from SJC - Landing/takeoff rates}
\begin{figure}[!ht]
\subfigure{\includegraphics[width=0.48 \linewidth]{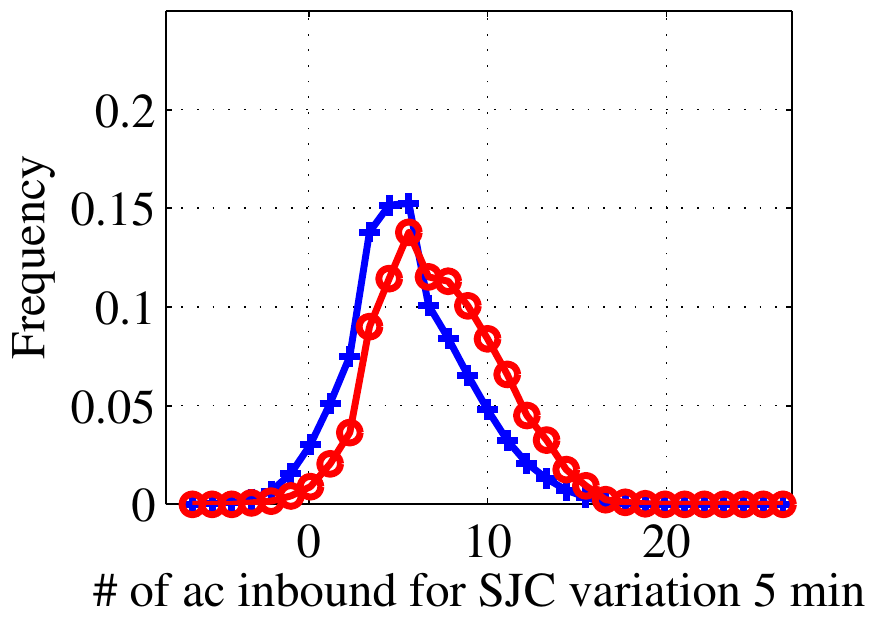}\label{fig:inboundSJC3}}
\subfigure{\includegraphics[width=0.48 \linewidth]{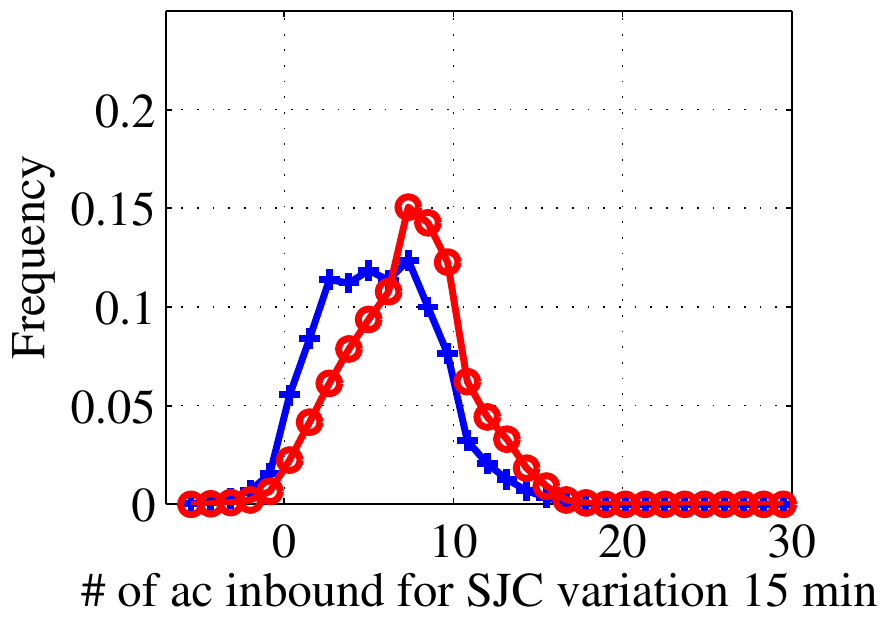}\label{fig:inboundSJC4}}
\caption{Analysis of the changes in number of aircraft incoming to SJC}\label{fig:inboundSJC}
\end{figure}

In terms of instantaneous or average number of aircraft, the distributions associated with the number of aircraft inbound to and outbound from SJC do not show significant differences. Figure~\ref{fig:inboundSJC} presents the distribution of the difference in the current number of inbound aircraft and the number of inbound aircraft 5 and 15 minutes ago. It appears an increase in the number of aircraft inbound for SJC tends to indicate an increase in the frequency of GAs at SFO. Note that the plots are not centered at 0, suggesting a correlation between the runway configuration used at SFO and the changes in traffic volume inbound for SJC.
The landing and takeoff rates at SJC over 5, 10 and 15 minutes are not presented; they do not appear to have a statistical impact on GAs at SFO.
%
\subsubsection{Number of aircraft on the airport surface, inbound}
\begin{figure}[!ht]
\subfigure{\includegraphics[width=0.48 \linewidth]{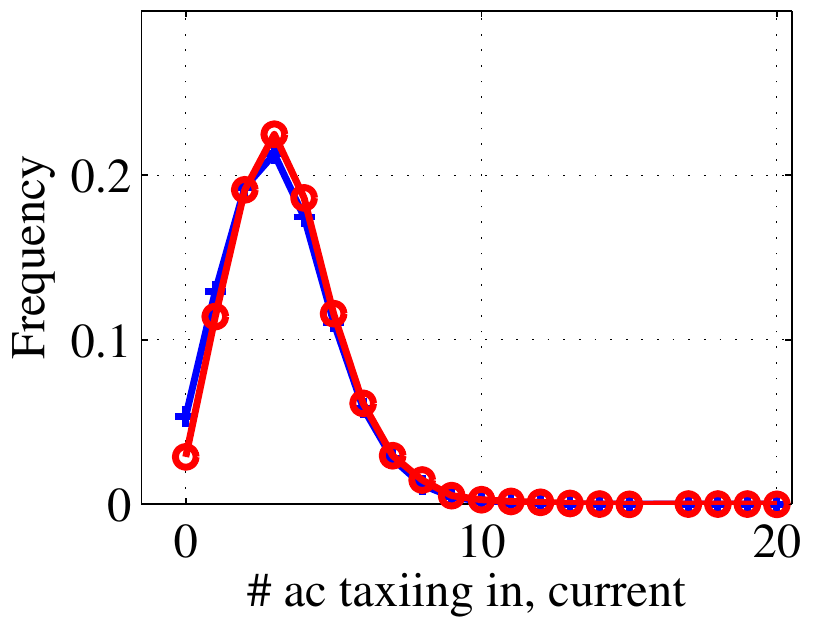}\label{fig:taxiInSFO1}}
\subfigure{\includegraphics[width=0.48 \linewidth]{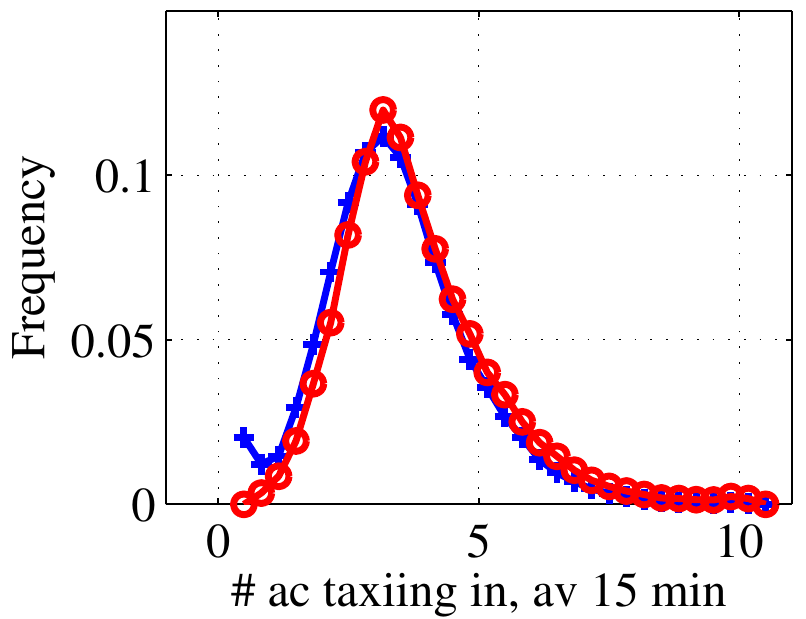}\label{fig:taxiInSFO2}}
\subfigure{\includegraphics[width=0.48 \linewidth]{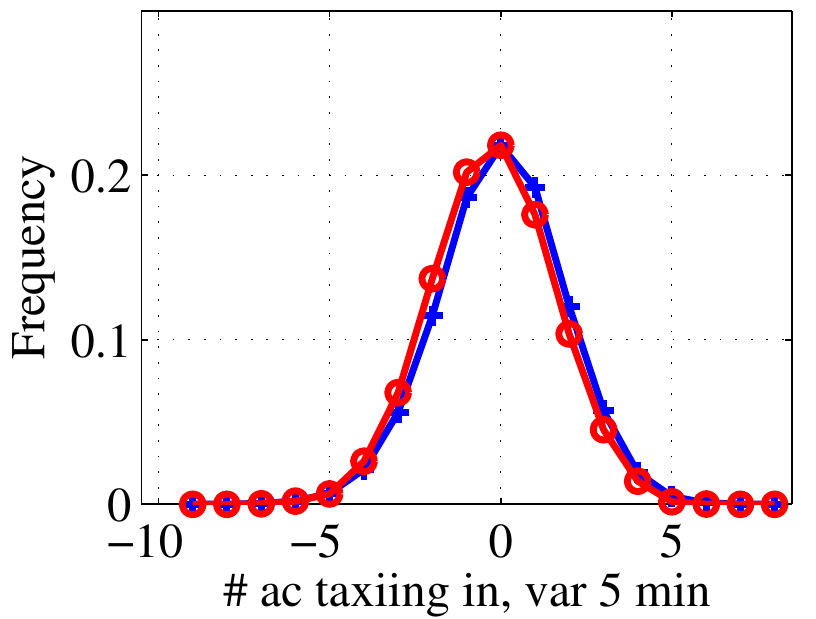}\label{fig:taxiInSFO3}}
\subfigure{\includegraphics[width=0.48 \linewidth]{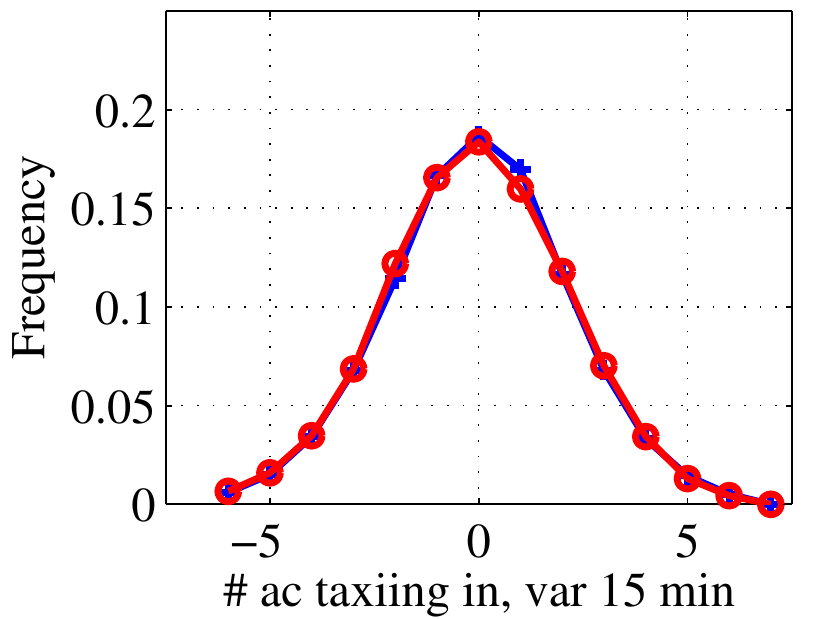}\label{fig:taxiInSFO4}}
\caption{Analysis of the number of aircraft taxiing at SFO, inbound}\label{fig:taxiinSFO}
\end{figure}

Figure~\ref{fig:taxiinSFO} presents the distributions associated with the number of inbound aircraft taxiing at SFO. Figure~\ref{fig:taxiInSFO1} shows the number of aircraft at the time the sample is taken and Figure~\ref{fig:taxiInSFO2} shows the average number of aircraft over the preceding 15 minutes. These measures  reflect the current congestion at the airport for aircraft taxiing-in. The shape of nominal and GA distributions are very similar, with the GA distributions slightly skewed toward higher aircraft counts.
 Figures~\ref{fig:taxiInSFO3}  and~\ref{fig:taxiInSFO4}  present distributions for the difference between the number of aircraft simultaneously present at the time of sample and 5 and 15 minutes ago, respectively. The plots would suggest that a high number of incoming aircraft slightly increases the probability of having a GA, but some GAs occur when there are only a few incoming aircraft.  The 5 minute variation in the number of inbound aircraft is slightly shifted toward the negative numbers for the GAs, meaning that GAs are more likely to occur when the number of aircraft inbound on the surface diminishes. This effect is not visible in the case of 15 minute variations.
\subsubsection{Number of aircraft on the airport surface, outbound}
\begin{figure}[!ht]
\subfigure{\includegraphics[width=0.48 \linewidth]{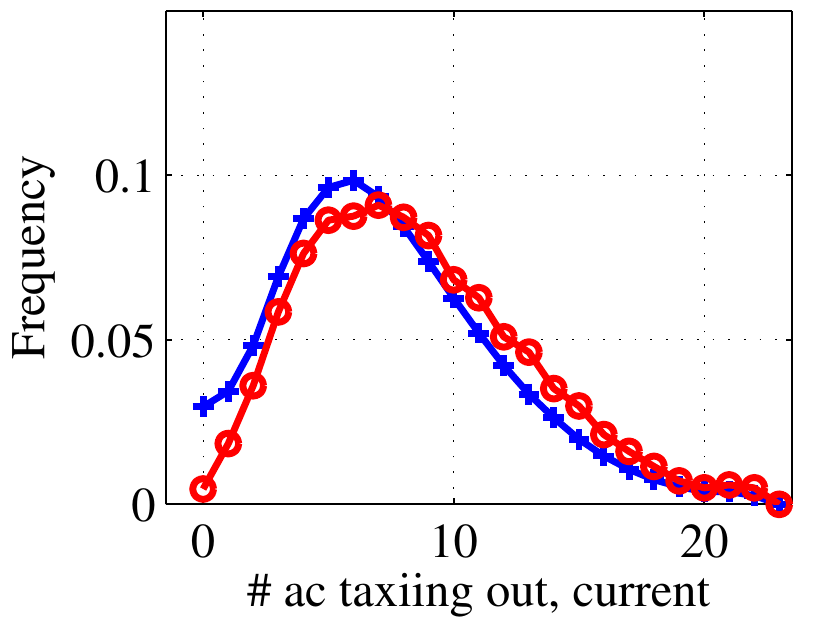}\label{fig:taxiOutSFO1}}
\subfigure{\includegraphics[width=0.48 \linewidth]{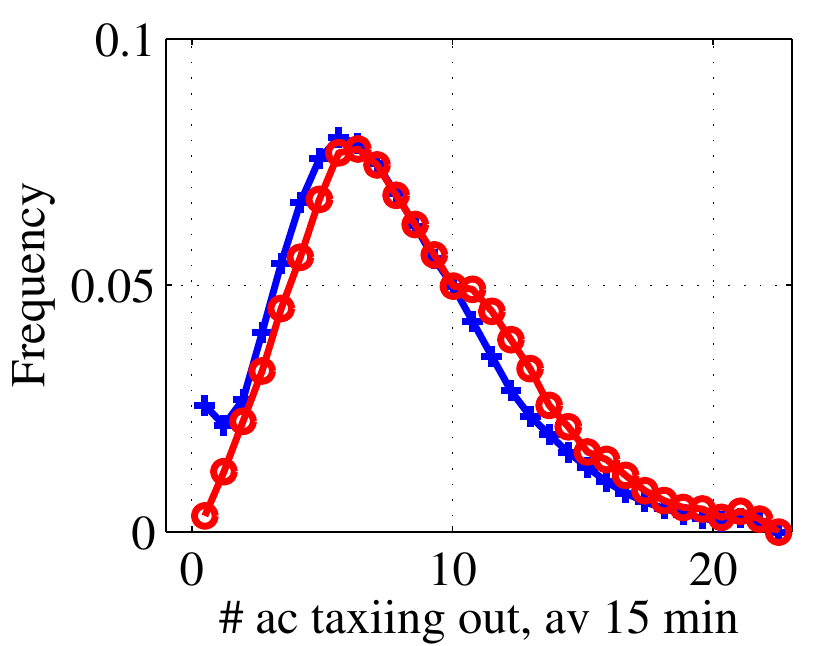}\label{fig:taxiOutSFO2}}
\caption{Analysis of the number of aircraft taxiing at SFO, outbound}\label{fig:taxiOutSFO}
\end{figure}

 Figure~\ref{fig:taxiOutSFO} presents the distributions of the number of inbound aircraft taxiing at SFO. Figure~\ref{fig:taxiOutSFO1} shows the number of aircraft at the time the sample is taken and Figure~\ref{fig:taxiOutSFO2} shows the average number of aircraft over the preceding 15 minutes. These measures reflect the current congestion at the airport, for aircraft taxiing-in. The shape of the distribution of the GAs and the nominal samples are very similar; there are slightly more GAs at the higher aircraft counts.
It appears that a high number of outbound aircraft has an impact on the probability of having a GA, but some GAs occur when there are only a few incoming aircraft. The 15 minutes plot suggests that having an average of more than 10 aircraft taxiing out has a significant impact on GAs.

The difference between the number of aircraft simultaneously present at the time of the sample and 5, 10 and 15 minutes in the past are not depicted, since they do not show any particularly interesting results. 
\subsubsection{Delays}
\begin{figure}[!ht]
\subfigure{\includegraphics[width=0.48 \linewidth]{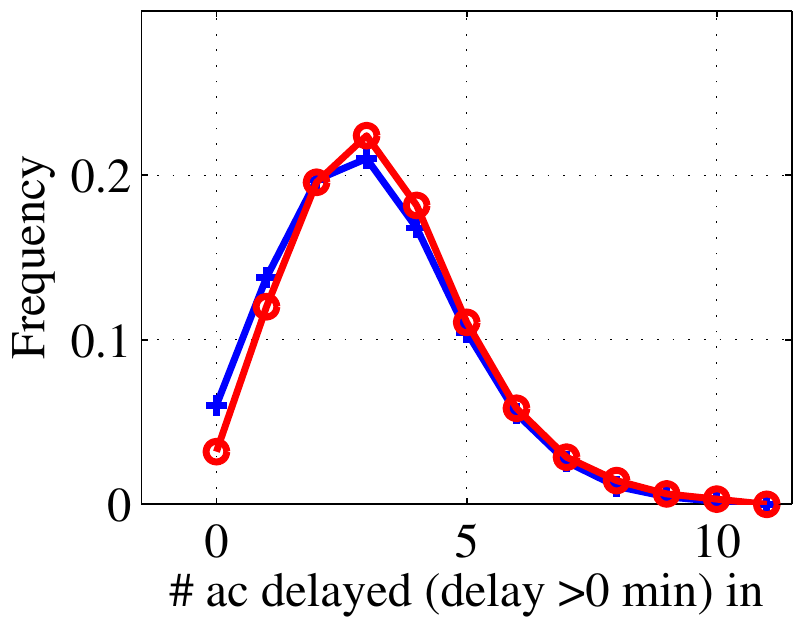}\label{fig:delay1}}
\subfigure{\includegraphics[width=0.48 \linewidth]{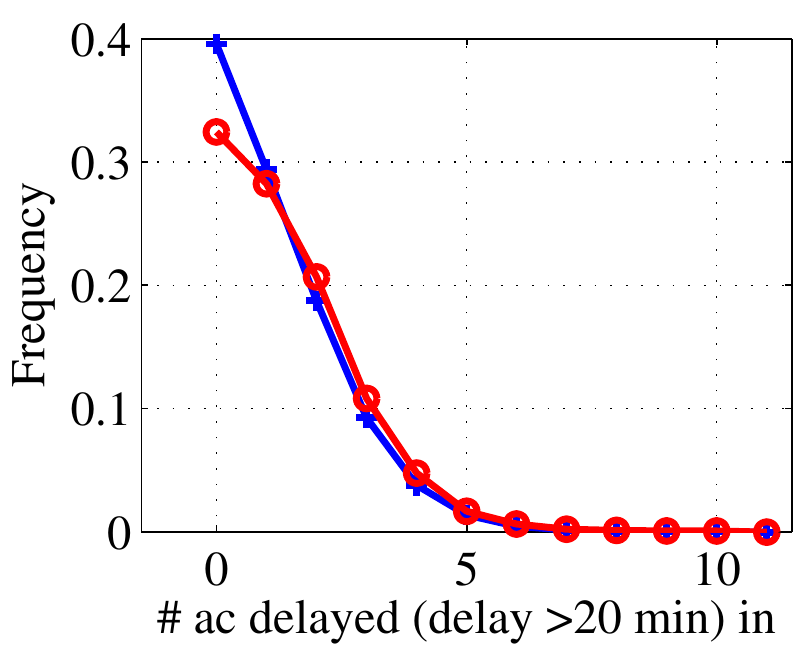}\label{fig:delay2}}
\subfigure{\includegraphics[width=0.48 \linewidth]{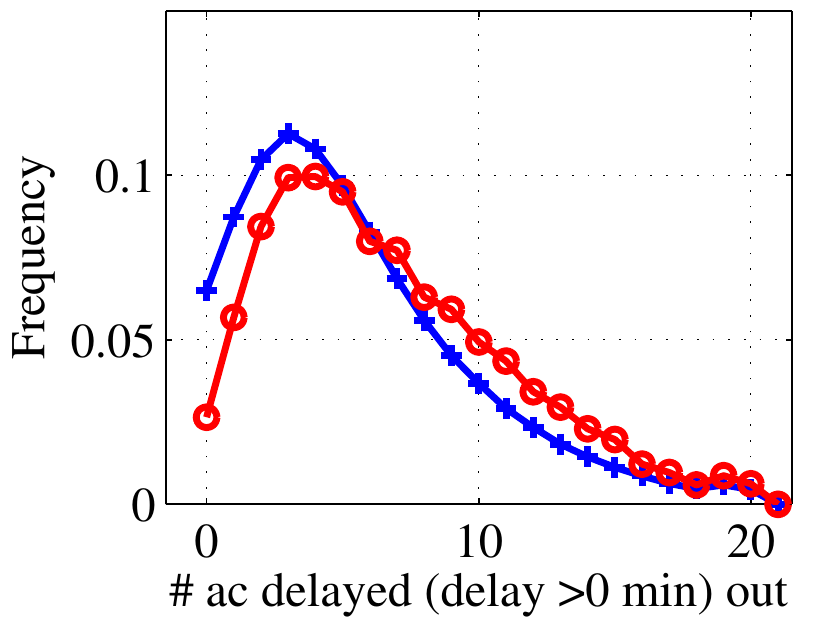}\label{fig:delay3}}
\subfigure{\includegraphics[width=0.48 \linewidth]{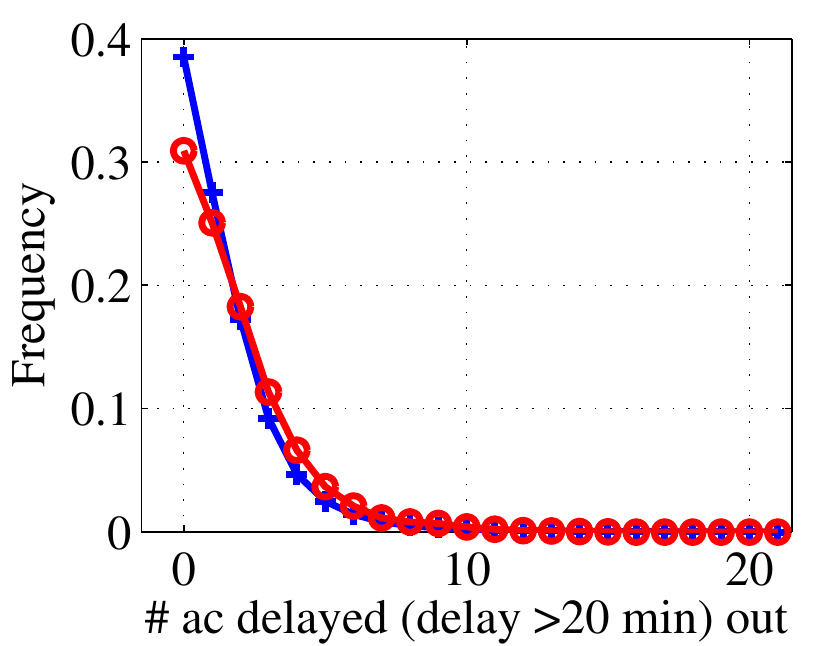}\label{fig:delay4}}
\caption{Analysis of the number of aircraft delayed at SFO}\label{fig:delay}
\end{figure}

Figure~\ref{fig:delay} shows the distributions as a function of the number of aircraft delayed. Figure~\ref{fig:delay1} shows the distribution of the number of inbound aircraft with a delay, taxiing in to the gate. Figure~\ref{fig:delay2}, presents the distribution for delays greater than 20 minutes. It appears that GAs are less likely to occur when there are no delayed aircraft taxiing to a gate. For delays greater than 20 minutes, although 40\% of the time there are no aircraft delayed to this extent, 33\% of the GAs occur under these conditions, indicating large delays may contribute to a GA. 

Figures~\ref{fig:delay3} and~\ref{fig:delay4} present the same distributions for the case of aircraft taxiing out. While 25\% of the time there are no more than two aircraft with a delay, 15\% of the GAs occur under these conditions. 

\subsubsection{Weather}
\begin{figure}[!ht]
\subfigure{\includegraphics[width=0.48 \linewidth]{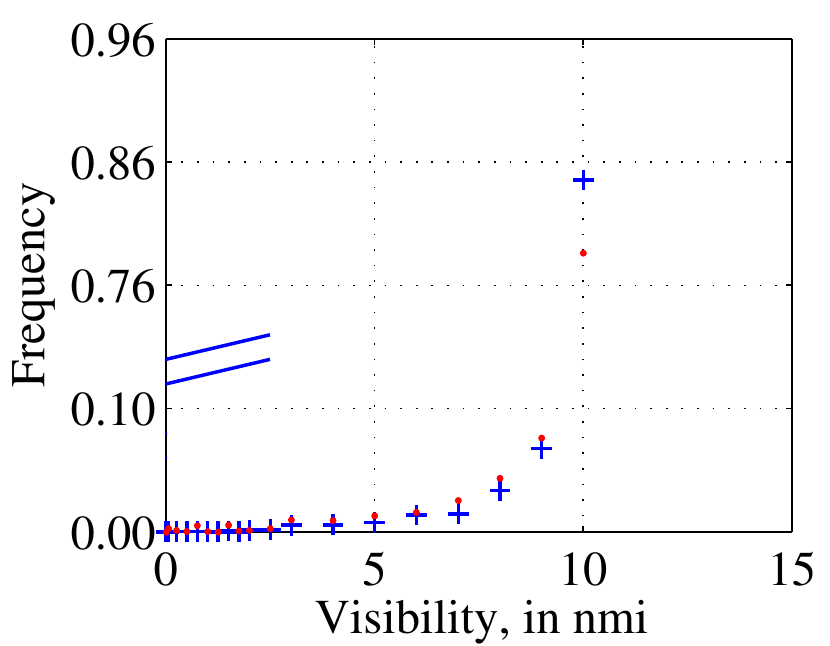}\label{fig:visibility}}
\subfigure{\includegraphics[width=0.48 \linewidth]{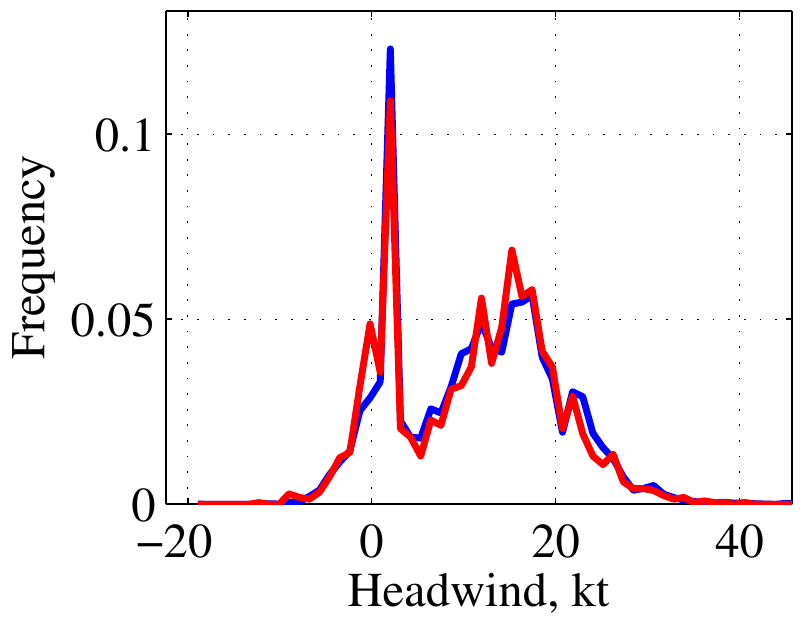}\label{fig:headwind}}
\subfigure{\includegraphics[width=0.48 \linewidth]{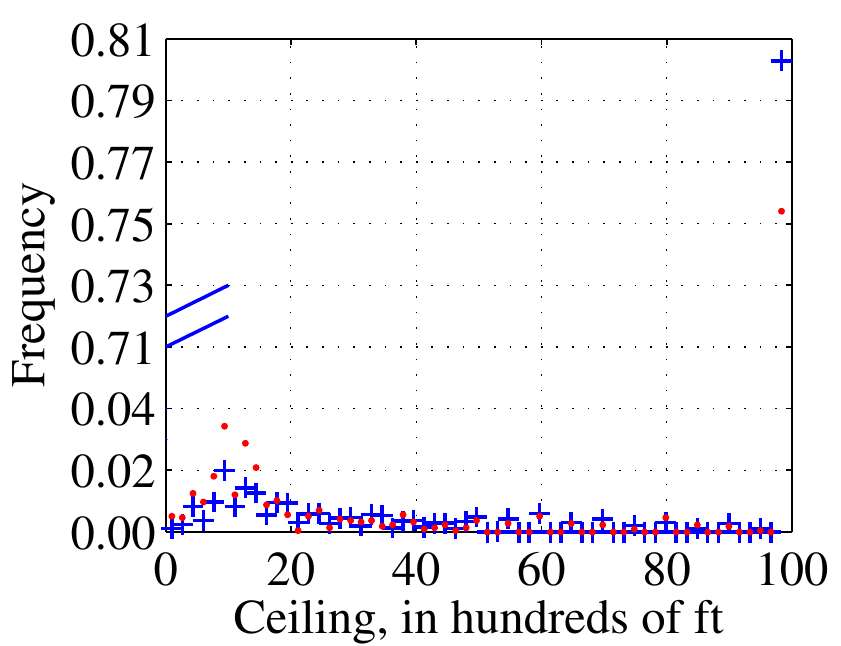}\label{fig:ceiling}}
\subfigure{\includegraphics[width=0.48 \linewidth]{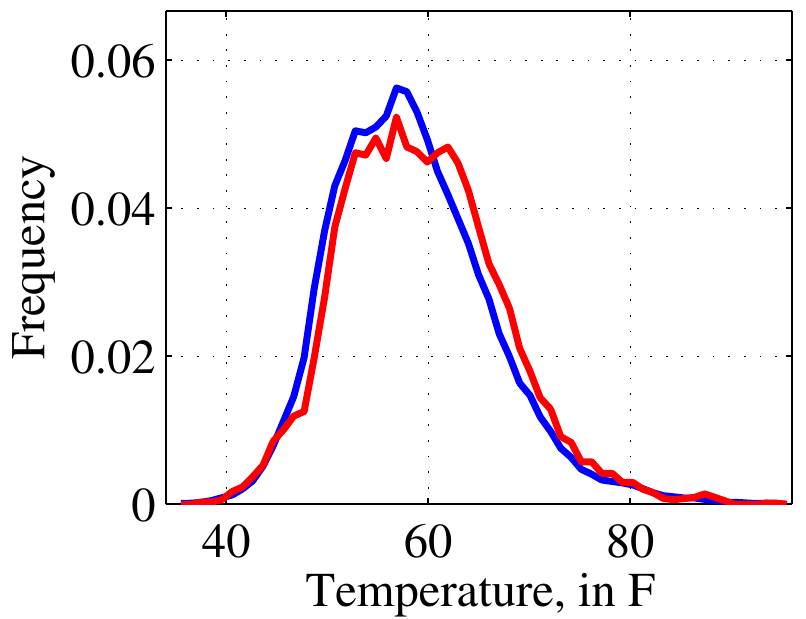}\label{fig:Temperature}}
\caption{Analysis of the weather related parameters}\label{fig:weather}
\end{figure}

Figure~\ref{fig:weather} shows the distribution of of nominal and GA flights as a function of various weather parameters. Figure~\ref{fig:visibility} presents the frequency of the samples as a function of the visibility. A visibility of 10 nmi indicates that the actual visibility was at least 10 nmi. It appears the visibility at SFO is greater than 10 nmi approximatively 83\% of the time, but only 75\% of go arounds occur during these conditions.  GAs appear to occur at a greater rate during low visibility conditions with 25\% of GAs occurring in 17\% of the time in which visibility is lower. Adverse weather condition significantly increase the probability of GA, only 25 \% of GA occur during poor weather conditions.
Figure~\ref{fig:headwind} presents the nominal and GA distributions as a function of headwind. A negative headwind corresponds to a tailwind. Most of the flights land with positive to no headwind, and the headwind does not appear to be a significant cause of GAs. In negative headwinds, that is tailwinds, GAs appear to be more frequent. The crosswind, is not depicted but does not seem to have an impact on GAs. The data suggests crosswinds are not a leading cause of GAs. 
Figure~\ref{fig:ceiling} presents the altitude of the sky's ceiling. All ceiling altitudes over 10,000 ft were trimmed to 10,000 ft. The data suggests a low ceiling is associated with an increase in the likelihood of a GA. Figure~\ref{fig:Temperature} shows the distributions as a function of temperature. The plots suggest GAs are more likely to occur at ``higher'' temperatures.

\subsection{Discussion of results}

From this analysis, three main factors leading to an increased probability of a GA are the weather, the airborne traffic density and aircraft mix, and finally, the ground traffic and its delays.   

\subsubsection{Weather:} When the visibility or the ceiling are low, the rate of GAs is much higher than in good weather conditions. A likely explanation is the lack of visibility at decision hight forcing the pilot to initiate a missed approach and return. Wind, including tailwind and crosswind do not appear to have a significant impact on the probability of GAs to occur. The weather has a direct impact on GAs, but at SFO, the number of GAs due to poor weather conditions is only 25 \%.

\subsubsection{Traffic density and aircraft mix:} The analysis suggests that having a large number of incoming aircraft increases the probability of having a GA. From a human factor's perspective, a large number of aircraft simultaneously present in the terminal airspace increases the workload of the controllers, probably leading to more ``operational errors'' and violation of landing minimum separation distance. The terminal airspace is rather small and congested, therefore dealing with many aircraft becomes quickly very complex and lead air traffic controllers to vector and reroute aircraft \cite{gariel11clustering}. In a previous study \cite{GarielDynamicIOmodel}, it was shown that limiting the number of aircraft simultaneously present in the TRACON tends to allow for more direct routes, hence reducing the perceived complexity, and eventually, maybe reducing the probability of a GA. However, there are some GAs that occur when there are only a few incoming aircraft, perhaps a testament to the sporadic and haphazard nature of the event. 

The aircraft mix appears to have an effect on the likelihood of GA. A high number of large aircraft and heavy aircraft landing in the past 5 to 15 minutes increases the probability of a GA. Possible explanations include a separation distance too small between aircraft.  

There is also evidence to suggest the variation in the number of planes being metered to OAK has an impact on the likelihood of a GA; a positive change in the number of aircraft incoming into OAK seems to increase the probability of a GA occurring at SFO. A possible explanation is again from a human factor's perspective for the controllers in charge of the sequencing and merging for SFO, OAK and SJC. Since most of the traffic in the TRACON is directed to SFO, a sudden increase in the number of aircraft inbound for OAK or SJC requires a shift of attention from the controller, probably breaking his current mental model \cite{reynolds2002structure} of the situation. This analysis shows the coupling effect between the airport, not only because of the traffic that need to be separated, but also from a controller's point of view. 

\subsubsection{Ground traffic and delays}
It appears that a large number of aircraft taxiing out at SFO increases the probability of a GA occurring. In addition, an average over 15 minutes of more than10 aircraft taxiing out has a visible impact. Note also that delays affecting either inbound or outbound aircraft increase the probability of GAs. It appears that human errors such as runway incursion, holding lines violation or late takeoff from runway are more likely to occur during high density outbound ground traffic and when delays affect ground traffic.


\section{An Alerting System for Go-Arounds}\label{sec:prediction}
In this section, we present a system to evaluate the potential of a GA. The first step is to classify GAs from nominal samples using the available data. The second step is to evaluate the potential of a GA at each time step. We first introduce the issues related to predicting these rare and poorly separated events, before presenting the results of our classification and prediction results based on the method of linear discriminant analysis. 
\subsection{Classification and Prediction issues}
There are a number of factors that make classification and temporal prediction of GAs difficult. First, GAs are a rare occurrence. Although this is auspicious form an air transportation perspective, it makes learning difficult and introduces a strong statistical bias in our corpus of training examples. To improve learning, it is natural to use a modified training set with roughly an equal number of nominal and GA samples. This can be accomplished by either withholding a large portion of nominal examples from the corpus, up-sampling the GAs, or a combination of the two methods. Unfortunately, these methods fail to address the more deep-seeded issue that GAs occur during standard phases of operation and therefore GA samples will generally have features closely aligned with nominal samples. In other words, nominal and GA samples are very poorly separated in the sample space. As false positives (nominal samples that are labelled as GA samples) are especially undesirable in our context, we must seek to learn relationships that separate the training data. Because of the poor separability of the samples, it is not possible to predict GAs and maintain a low rate of false positives. Therefore, we aim at evaluating ``high risk'' time samples which have a higher probability of having a GA. These high risk time samples will be denoted by an ``alert level''. The objective is to maximize the number of GAs positively identified during the alert level while minimizing the total number of samples in the alert level. 

%
%
%
\subsection{Linear discriminant analysis}
Linear discriminant analysis is a statistical method commonly used to separate samples into several classes~\cite{Bishop07}. In our case, we are concerned with only two classes of samples: nominal state vector samples with label $y=0$ and GA state samples with label $y=1$. We will assume GAs and nominal samples are generated according to a two-label Gaussian mixture model. For this purpose, our corpus of samples, $\{x_i\}_{i=1}^{N}$, $x_i \in \mathbb{R}^{135}$ is split into two groups, a training group and a test group. 
A strong assumption made by LDA is that the conditional probability density functions $p( x | y=0)$ and $p( x|y=1)$ are both normally distributed with mean and covariance $\left( \mu_0, \Sigma_{0}\right)$ and $\left( \mu_1, \Sigma_{1}\right)$, respectively. Then, a feature vector $x$ is assigned the label 0 if it satisfies \begin{equation}\label{eq:LDA}
\begin{split}
 ( x-\mu_0)^T &\Sigma_{y=0}^{-1}( x- \mu_0)+\mathrm{ln}|\Sigma_{y=0}|-
 \\
& ( x- \mu_1)^T \Sigma_{y=1}^{-1} ( x- \mu_1)-\mathrm{ln}|\Sigma_{y=1}|<T,
 \end{split}
 \end{equation} where $T$ is a threshold value that reflects the frequency of a label and the $|\Sigma|$ denotes the determinant of $\Sigma$. Otherwise, $x$ is assigned a value of 1. By sweeping $T$ downward from a very large number to zero, we can progressively decrease the rate at which 0 labels are assigned. That is, we can control the number of GAs that we are able to identify correctly. However, correctly identifying the bulk of the GAs comes at the expense of having a large number of false positives.  

 \subsection{An alert system for GA}
\begin{figure}[htbp]
\includegraphics[width = 0.96\linewidth]{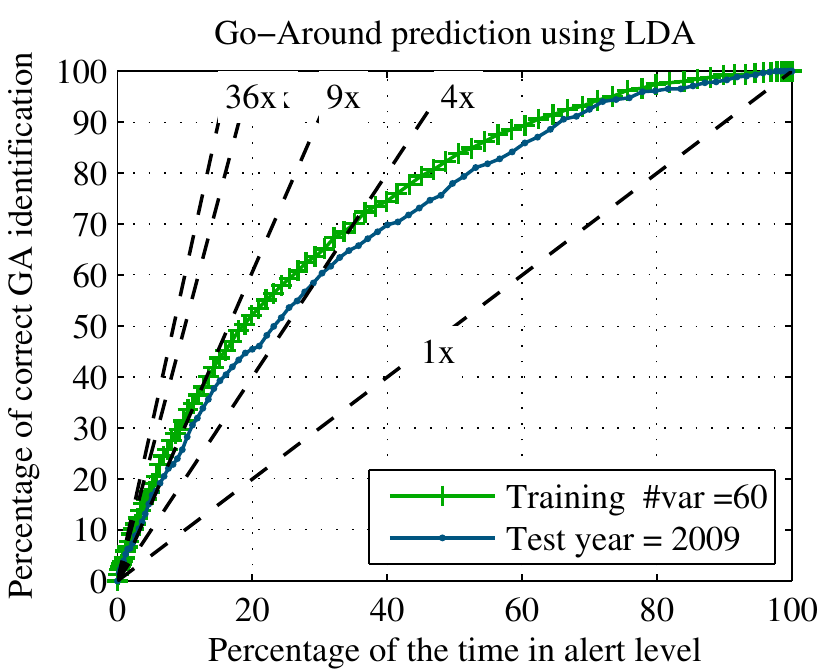}
\caption{Classification of GA using LDA}\label{fig:LDA}
\end{figure}

\begin{figure}[htbp]
\includegraphics[width = 0.96\linewidth]{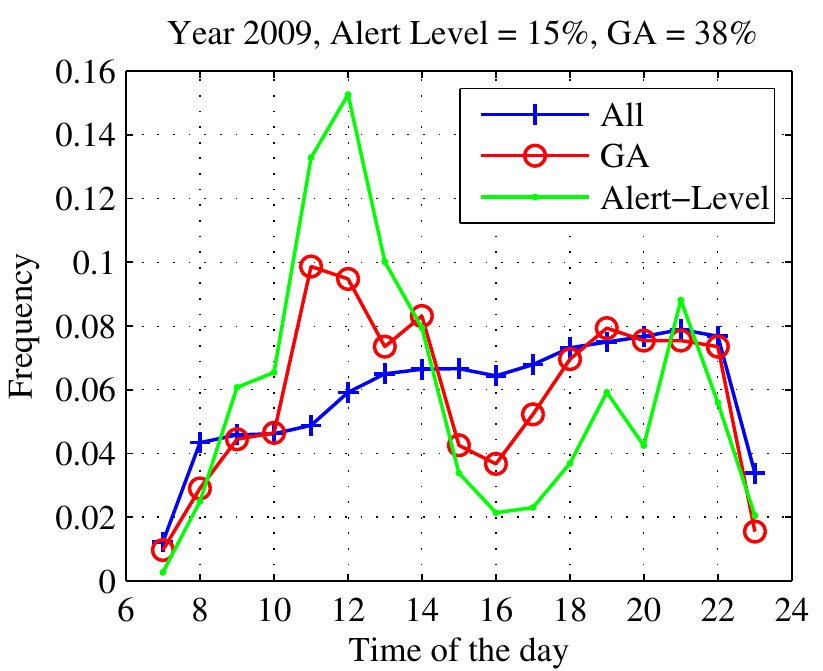}
\caption{Time distribution of nominal samples, GA and alert-level}\label{fig:timehist}
\end{figure}

 We used LDA on three years of data to classify GA and nominal samples. A separate year of data was withheld to test the classifier's predicative capabilities. The training set consists of randomly selected nominal samples and all the GA from 2006 to 2008. The test set contains all the available samples from 2009.  Figure \ref{fig:LDA} presents the result of the classification and prediction. The green curve (top) is corresponds to the training set and the blue curve (bottom) to the test set. The horizontal-axis represents the fraction of time predicted as alert-level, that is the fraction of the time where GAs are likely to occur. The $y$-axis represents the proportion of actual GAs that occurred during a period of alert-level. 
The crosses on the green line correspond to the different values of the threshold $T$ (Eq. \ref{eq:LDA}). By fixing a value of $T$, one can choose the amount of samples in alert-level.  The dashed black line indicate the increased probability of a GA to occur during alert-level compared to the remaining of the time. The line 1x indicates complete randomness of the prediction. Then, when the green line is over the 4x line, a GA is 4 times more likely to occur than during the remaining of the time. For instance, the green line intercepts the \textit{9x} line at 15\% of the time in threat level. This means that the predictor can be in ``threat level'' 15\% of the time and capture 39\% of the GAs. During alert-level time, GAs are 9 times more likely to occur than during the remaining 85\% of the time.
To further analyze the results of the predictive system, we looked at the distribution over time of the samples predicted in alert level.  Figure~\ref{fig:timehist} presents the time distribution of the nominal samples, GA and samples identified in alert-level for 2009. In this example, the value of $T$ was selected so that 15\% of the samples are in threat level, capturing 39\% of the GA. The alert-level curve follows the same pattern as the GA  over time. The predictive system slightly over-estimate the risk during the morning peak and under-estimates the risks during the evening peak. 

\section{Conclusion}\label{sec:conclusions}
This paper investigated a number of airport operational features, each of which is readily accessible to on-duty air traffic controllers, and the role fluctuations in them may play in precipitating a missed approach. We showed the important interconnection between surface operations, airborne operations, airports located in the vicinity of each other, air traffic control and delays. By analyzing how the distribution of these features varied between nominal and go-around operations, we provided a statistical mechanism to gain insight into which factors are more likely to be a discernible precursor to go-arounds. Interpretations for these results were provided in terms of the current operational policies in place at busy metropolitan airports. Armed with the new insight afforded by these statistics, we proposed a framework for developing an automated alert system for high missed approach potential situations. Although the machine learning techniques employed for this purpose would mandate the alert system be ``on" for an large amount of time in order to capture most of the go-arounds. It appears that the prediction of go-arounds is a very challenging task and this study highlighted unexpected factors that have an impact on the probability of a missed approach. These factors such as airport coupling effect for controllers need to be accounted for in the future design of metroplexes and high-density operations.
\addtolength{\baselineskip}{-0.1\baselineskip} 

\bibliographystyle{IEEEtran}
\bibliography{main} 
  
  \section*{Biographies}
\textbf{Maxime Gariel} is a post-doctoral associate in the Laboratory for Information and Decision System at the Massachusetts Institute of Technology. He obtained the Engineering degree in aerospace in 2008 from ISAE-Supaero, France. He received a dual M.S in aerospace engineering in 2008 from the Georgia Institute of Technology and ISAE-Supaero. He obtained his Ph.D. in aerospace engineering from Georgia-Tech in 2010. His research interests include air transportation, avionics, safety, data-mining and its applications to air transportation.

\textbf{Kevin Spieser} received the B.A.Sc. degree and M.A.Sc. degrees, both in electrical engineering from the University of Waterloo in 2006 and 2008, respectively. He is currently a Ph.D. candidate in the Department of Aeronautics and Astronautics and MIT. His main research interests are in the area of multi-agent systems and their application to the study of crowd control, traffic regulation, and air transportation systems.

\textbf{Emilio Frazzoli} is a Professor of Aeronautics and Astronautics with the Laboratory for Information and Decision Systems at the Massachusetts Institute of Technology. He received a Laurea degree in Aerospace Engineering from the University of Rome, "Sapienza" , Italy, in 1994, and a Ph. D. degree in Navigation and Control Systems from the Department of Aeronautics and Astronautics of the Massachusetts Institute of Technology, in 2001. He was the recipient of a NSF CAREER award in 2002. He is an Associate Fellow of the AIAA and a Senior Member of the IEEE.   
\end{document}